# Primordial black hole formation and multimessenger signals in a complex singlet extension of the standard model

Fa Peng Huang, Chikako Idegawa,* and Aidi Yang

*MOE Key Laboratory of TianQin Mission, TianQin Research Center for Gravitational Physics and School of Physics and Astronomy, Frontiers Science Center for TianQin, Gravitational Wave Research Center of CNSA, Sun Yat-sen University (Zhuhai Campus), Zhuhai 519082, China*

(Received 6 November 2025; accepted 13 February 2026; published 4 March 2026)

We investigate the formation of primordial black holes (PBHs) induced by a first-order electroweak phase transition in a realistic renormalizable framework, the complex singlet extension of the Standard Model. We perform a quantitative analysis of the PBH abundance and identify parameter regions consistent with current microlensing constraints. Furthermore, we show that the same parameter space predicts observable stochastic gravitational waves within the sensitivities of future space-based detectors, as well as a sizable deviation in the Higgs triple coupling that can be probed at future lepton colliders. Our results highlight a comprehensive multimessenger framework in which PBHs, gravitational waves, and collider observations can jointly test the dynamics of a strongly first-order electroweak phase transition in the early Universe.

## I. INTRODUCTION

Primordial black holes (PBHs) are hypothetical black holes formed in the early Universe, long before the formation of stars and galaxies [1–3]. They can serve as unique probes of the early Universe, providing insights into high-energy physics phenomena that cannot be tested in laboratory experiments. A wide variety of mechanisms for the formation of PBHs have been proposed. Over the past few decades, one extensively studied class of models involves scenarios where large primordial quantum fluctuations are generated at small scales in some inflation models [4–7]. In such frameworks, the enhanced curvature perturbations can collapse gravitationally upon horizon reentry, giving rise to PBHs [3]. More recently, increasing attention has been devoted to alternative mechanisms in which PBHs are produced during cosmological first-order phase transitions in the early Universe. In these scenarios, the specified dynamical processes associated with the phase transition can generate regions of overdensity that may gravitationally collapse into PBHs [8–23]. These studies open up a new window for connecting PBH formation with the microphysics of the early Universe, providing a complementary probe of new physics models.

Among the various mechanisms proposed for PBH formation during a strong first-order phase transition, a particularly well-motivated scenario involves a nonuniform nucleation rate in time. Certain regions of the Universe can undergo bubble nucleation earlier or later than others. During such a transition, true-vacuum bubbles nucleate stochastically through thermal tunneling, and different Hubble patches complete the transition at slightly different times [8,24]. As a result, regions where bubble nucleation occurs later, referred to as delayed patches, remain in the false vacuum longer and thus retain a larger vacuum energy density. Since the false-vacuum energy does not redshift while the radiation energy density decreases as the Universe expands, the delayed-decayed regions gradually become overdense compared to the surrounding space [9,25]. Once the energy-density contrast between the normal-decayed and delayed-decayed regions exceeds a critical threshold, the region collapses into a PBH upon horizon reentry. The PBH fraction $f_{\rm PBH}$ is determined by the rare probability that a Hubble patch remains in the false vacuum up to such a delayed time, making $f_{\rm PBH}$ extremely sensitive to the detailed dynamics of the phase transition [15,18]. This delayed vacuum decay mechanism therefore connects the microscopic physics of the phase transition to macroscopic cosmological observables.

While the possibility of PBH formation from a first-order phase transition has been studied extensively, most of the existing analyses have been performed in a model-independent framework or within phenomenological toy models that parametrize the thermal potential. In particular, studies focusing on the electroweak phase transition (EWPT) have often adopted such approaches, since the properties of the EWPT, such as its strength and critical temperature, depend sensitively on the underlying model [26–28], and a full numerical treatment of the finite-temperature potential is computationally demanding. As a result, the dynamics of PBH formation have typically been discussed in a general manner without specifying a concrete model. A few studies have explored PBH formation within specific particle-physics models such as Refs. [14,15,18,19,21–23,29–32]. In this work, we instead focus on a realistic renormalizable setup, a complex singlet extension of the Standard Model (CxSM) [33–48], and examine the possibility of PBH formation in this framework, together with its phenomenological implications for gravitational waves (GWs) and collider observables.

*Contact author: idegawa@mail.sysu.edu.cn

As a well-motivated and minimal extension of the Standard Model (SM), the CxSM introduces a gauge-singlet complex scalar field $S$ that couples to the Higgs doublet. This model simultaneously accommodates a viable dark matter (DM) candidate and allows for a strong first-order EWPT [42,43]. In particular, we focus on the so-called degenerate-scalar scenario, in which the additional scalar has a mass nearly degenerate with the observed Higgs boson [49,50]. This setup is known to evade current DM direct-detection bounds [51] due to destructive interference between multiple scalar-mediated amplitudes, while remaining consistent with collider constraints [52,53].

The existence of DM provides clear evidence for physics beyond the SM; however, no conclusive DM signal has been observed so far. While this null result is often interpreted in terms of decoupling scenarios, such as very heavy new states or extremely suppressed couplings to the SM particles, the steadily improving experimental sensitivity without corresponding discoveries motivates the exploration of qualitatively different possibilities. In contrast to such decoupling-based interpretations, the degenerate-scalar realization of the CxSM provides a qualitatively different possibility: the nonobservation of DM can be naturally understood without invoking decoupling from the SM.

It should also be emphasized that, in the parameter region relevant for a strong first-order EWPT in the CxSM, the relic abundance of the pseudoscalar DM component is typically much smaller than the observed DM density. This feature has been explicitly demonstrated in Ref. [42], where it was shown that the pseudoscalar DM in the degenerate-scalar scenario cannot account for the full dark matter abundance. A concrete realization of a two-component DM scenario within the CxSM, in which the missing relic density is supplemented by an additional fermion DM candidate, was subsequently studied in Ref. [47]. Therefore, in the present work, we do not require the pseudoscalar component to explain the entire DM abundance. While PBHs can in principle contribute to the DM abundance, we do not perform a detailed quantitative assessment of their contribution in this work. Instead, PBH formation is mainly used as a cosmological probe of the dynamics of the EWPT. We note, however, that PBHs could potentially complement other DM components and jointly account for the observed DM relic abundance.

In this work, we investigate the possibility of PBH formation induced by a first-order EWPT in the CxSM. We explore how the dynamics of the EWPT, determined by the scalar potential parameters, affect the resulting PBH abundance, and we compare the predicted fractions with recent microlensing constraints from HSC and OGLE [54–56]. In addition to PBH production, we also study two complementary phenomenological consequences of a strong first-order EWPT: the generation of stochastic GWs [57,58], and the deviation of the Higgs triple coupling from its SM prediction [59,60]. Through this multimessenger approach, we demonstrate that the same underlying phase transition can simultaneously lead to observable signals in cosmology, GW astronomy, and collider experiments.

This paper is organized as follows: In Sec. II, we review the formalism of PBH formation arising from a first-order EWPT. In Sec. III, we introduce the CxSM, define the relevant parameters, and discuss the degenerate-scalar scenario. The numerical results for PBH formation are presented in Sec. IV, where we analyze how the model parameters affect the transition dynamics and compare the predicted PBH abundances with current microlensing constraints from HSC and OGLE. In Sec. V, we investigate the GW signatures of the EWPT and their correlations with collider observables such as the Higgs triple coupling deviation. Finally, Sec. VI summarizes our findings and discusses the implications of this comprehensive multimessenger framework for probing the dynamics of the EWPT.

## II. PRIMORDIAL BLACK HOLE FORMATION FROM ELECTROWEAK PHASE TRANSITION

In various extensions of the SM, a first-order EWPT plays an important role in the early Universe, such as those that lead to GW production and PBH formation. In particular, a first-order EWPT is often referred to as a strong first-order when the sphaleron process becomes inefficient inside the broken phase, which is approximately characterized by

$$\frac{v_C}{T_C} > 1. \tag{1}$$

$T_C$ denotes the critical temperature at which the effective potential develops two degenerate minima, and $v_C$ is the vacuum expectation value (VEV) of the Higgs doublet at $T_C$. We here outline the procedure used to evaluate the PBH abundance generated from such a transition, following the framework developed in Refs. [15,18,22,61]. First, the decay rate of the false vacuum is given by [62]

$$\Gamma(T) \sim T^4 \left(\frac{S(T)}{2\pi}\right)^{3/2} e^{-S(T)}, \tag{2}$$

where $S(T) = S_3(T)/T$, and $S_3(T)$ is the three-dimensional Euclidean action of a critical bubble. The nucleation temperature $T_N$ is defined by the condition

$$\Gamma(T_N) H^{-4}(T_N) = 1. \tag{3}$$

The spatial average of the false vacuum fraction at time $t$ is given by

$$F(t) = \exp\left[-\frac{4\pi}{3}\int_{t_R}^{t} dt'\, \Gamma(t') a^3(t') r^3(t,t')\right], \tag{4}$$

where $r(t,t')$ denotes the comoving radius of the true vacuum bubble, defined as

$$r(t,t') \equiv \int_{t'}^{t} \frac{d\tilde{t}}{a(\tilde{t})}. \tag{5}$$

The time variable can be converted into temperature using the Hubble parameter through

$$\frac{dt}{dT} = -\frac{1}{TH(T)}. \tag{6}$$

Then, we obtain

$$a(t')r(t,t') = a(T')\int_{T}^{T'} \frac{d\tilde{T}}{T' H(\tilde{T}) a(T')} = \frac{1}{T'}\int_{T}^{T'} \frac{d\tilde{T}}{H(\tilde{T})}. \tag{7}$$

Accordingly, the false vacuum fraction can be rewritten as a function of temperature:

$$F(T) = \exp\left[-\frac{4\pi}{3}\int_{T}^{T_R} \frac{dT'\Gamma(T')}{T'H(T')} a^3(T') r^3(T,T')\right], \tag{8}$$

where $T_R$ is the reference temperature that depends on the relevant region. The initial value of $F(T)$ is set differently for the normal-decayed and the delayed-decayed regions [63]:

$$F(T) = \exp\left[-12\pi(M_{\rm Pl}\xi_g)^4 \int_{T}^{T_C} \frac{dT'\Gamma(T')}{T'^6}\left(\frac{1}{T}-\frac{1}{T'}\right)^3\right], \quad \text{(normal-decayed region)}, \tag{9}$$

$$F(T) = \exp\left[-12\pi(M_{\rm Pl}\xi_g)^4 \int_{T}^{T_d} \frac{dT'\Gamma(T')}{T'^6}\left(\frac{1}{T}-\frac{1}{T'}\right)^3\right], \quad \text{(delayed-decayed region)}, \tag{10}$$

where $M_{\rm Pl} = 2.435 \times 10^{18}$ GeV and $\xi_g = \sqrt{30/(\pi^2 g)}$, with $g = 108.75$. For the normal-decayed region, we use the critical temperature $T_C$ as the reference temperature, while $T_d$, corresponding to the delayed-decayed region, is determined later to satisfy the PBH formation condition.

During a first-order EWPT, the Universe contains regions remaining in the false vacuum and regions that have transitioned to the true vacuum. The difference between the two regions is given by

$$\Delta V = V_{\rm false} - V_{\rm true}. \tag{11}$$

The total energy density of the Universe can be expressed as

$$\rho(T) = F(T) V_{\rm false} + (1 - F(T)) V_{\rm true}. \tag{12}$$

Choosing $V_{\rm true} = 0$ as the reference, the vacuum energy density becomes

$$\rho_v(T) = F(T)\Delta V. \tag{13}$$

The evolution of the radiation energy density is then described by

$$\frac{d\rho_r(T)}{dT} = \frac{4\rho_r(T)}{T} - \frac{d\rho_v(T)}{dT}, \tag{14}$$

and the Hubble parameter is determined by

$$H^2 = \left(\frac{\dot{a}}{a}\right)^2 = \frac{1}{3M_{\rm Pl}^2}(\rho_v(T) + \rho_r(T)), \tag{15}$$

where $\dot{a}$ denotes the derivative of the scale factor with respect to the temperature $T$. By substituting the above energy densities into Eq. (15), we can iteratively solve for the Hubble parameter and the scale factor until convergence.

For PBH formation, the relevant quantity is the contrast in the total energy density between the delayed- and normal-decayed regions:

$$\delta \equiv \frac{\rho_{\rm delay}}{\rho_{\rm normal}} - 1 = \frac{\rho_{r,\rm delay} + \rho_{v,\rm delay}}{\rho_{r,\rm normal} + \rho_{v,\rm normal}} - 1. \tag{16}$$

In this work, we adopt $\delta_c = 0.45$ as the threshold value for PBH formation [64], such that $\delta \geq \delta_c$ must be satisfied.[1] The temperature $T_d$ for the delayed region is chosen such that the peak of $\delta$ in Eq. (16) reaches 0.45. The temperature $T_d$ is lower than the nucleation temperature $T_N$, but if it becomes much lower, the probability of remaining in the false vacuum state rapidly decreases.

[1] For studies discussing the threshold value $\delta_c$, see, for example, Refs. [22,65–67].

The resulting PBH abundance produced by the EWPT is given by [15]

$$f_{\rm PBH}^{\rm EW} \simeq 1.49\times 10^{11}\left(\frac{0.25}{\Omega_{\rm CDM}}\right)\left(\frac{T_{\rm PBH}}{100\ {\rm GeV}}\right)P(t_N), \quad (17)$$

where

$$P(t_N) = \exp\left[-\frac{4\pi}{3}\int_{t_c}^{t_d}\frac{a_{\rm delay}^3(t)}{a_{\rm delay}^3(t_{\rm PBH})}\frac{1}{H_{\rm delay}^3(t_{\rm PBH})}\Gamma_{\rm delay}(t)dt\right]. \quad (18)$$

Here, $\Omega_{\rm CDM}$ denotes the present cold DM density normalized by the total energy density, and $T_{\rm PBH}$ ($t_{\rm PBH}$) represents the temperature (time) at PBH formation. The PBH mass is evaluated as

$$M_{\rm PBH} = \gamma\frac{4\pi}{3}H_{\rm delay}^{-3}(t_{\rm PBH})(\rho_{r,{\rm delay}} + \rho_{v,{\rm delay}}), \quad (19)$$

where $\gamma$ denotes the fraction of the horizon mass that collapses into a PBH, and we adopt $\gamma = 0.2$ in our analysis [3,18,68].

As mentioned in Sec. I, previous studies on PBH formation from the EWPT have often been carried out in a model-independent framework or by employing phenomenological toy models that parametrize the thermal potential. In this work, we instead focus on a well-motivated renormalizable extension of the SM—namely, the CxSM—and investigate the possibility of PBH formation, as well as the testability of the relevant parameters through future GW and collider experiments.

## III. THE COMPLEX SINGLET EXTENDED HIGGS MODEL

### A. Model introduction

The CxSM is an extension of the SM that introduces a complex SU(2) gauge-singlet scalar field [33]. In our study, we consider the following scalar potential:

$$V_0(\Phi, S) = \frac{m^2}{2}\Phi^\dagger\Phi + \frac{\lambda}{4}(\Phi^\dagger\Phi)^2 + \frac{\delta_2}{2}\Phi^\dagger\Phi|S|^2 + \frac{b_2}{2}|S|^2 + \frac{d_2}{4}|S|^4 + \left(a_1 S + \frac{b_1}{4}S^2 + {\rm H.c.}\right), \quad (20)$$

where a global U(1) symmetry of $S$ is softly broken by the $a_1$ and $b_1$ terms. In the following, all the couplings in Eq. (20) are assumed to be real. When the linear term in $S$ is absent, the potential possesses a $Z_2$ symmetry ($S \to -S$). If the singlet $S$ acquires a VEV, this $Z_2$ symmetry is spontaneously broken, leading to the domain-wall problem [49]. To avoid this issue, we include the linear term of $S$ in the potential Eq. (20), which explicitly breaks the $Z_2$ symmetry and thus prevents the formation of domain walls. Although several U(1)-breaking terms are allowed in the potential, not all of them are necessary to achieve a strong first-order EWPT and viable DM. Therefore, we adopt a minimal set of operators that is closed under renormalization.

We parametrize the scalar fields as

$$\Phi = \begin{pmatrix} G^+ \\ \frac{1}{\sqrt{2}}(v + h + iG^0) \end{pmatrix}, \quad (21)$$

$$S = \frac{1}{\sqrt{2}}(v_S + s + i\chi), \quad (22)$$

where $v(\simeq 246.22\ {\rm GeV})$ and $v_S$ denote the VEVs of $\Phi$ and $S$, respectively. The Nambu-Goldstone bosons $G^+$ and $G^0$ are absorbed by the $W$ and $Z$ bosons, respectively, after the electroweak symmetry breaking. Since we assume that all parameters in Eq. (20) are real, the scalar potential is invariant under the $CP$ transformation ($S \to S^*$). As a result, the real and imaginary components of $S$ do not mix, and the stability of $\chi$ is ensured, making it a scalar DM candidate.

The first derivatives of $V_0$ with respect to $h$ and $s$ are, respectively, given by

$$\frac{1}{v}\left\langle\frac{\partial V_0}{\partial h}\right\rangle = \frac{m^2}{2} + \frac{\lambda}{4}v^2 + \frac{\delta_2}{4}v_S^2 = 0, \quad (23)$$

$$\frac{1}{v_S}\left\langle\frac{\partial V_0}{\partial s}\right\rangle = \frac{b_2}{2} + \frac{\delta_2}{4}v^2 + \frac{d_2}{4}v_S^2 + \frac{\sqrt{2}a_1}{v_S} + \frac{b_1}{2} = 0, \quad (24)$$

where $\langle\cdots\rangle$ indicates that all fluctuation fields are set to zero. Note that a nonzero value of $v_S$ is enforced by $a_1 \neq 0$.

The mass matrix of ($h$, $s$) is expressed as

$$\mathcal{M}_S^2 = \begin{pmatrix} \lambda v^2/2 & \delta_2 v v_S/2 \\ \delta_2 v v_S/2 & \Lambda^2 \end{pmatrix}, \qquad \Lambda^2 \equiv \frac{d_2}{2}v_S^2 - \sqrt{2}\frac{a_1}{v_S}. \quad (25)$$

The mass matrix in Eq. (25) is diagonalized by an orthogonal matrix $O(\theta)$ as

$$O(\theta)^\top \mathcal{M}_S^2 O(\theta) = \begin{pmatrix} m_{h_1}^2 & 0 \\ 0 & m_{h_2}^2 \end{pmatrix},$$

$$O(\theta) = \begin{pmatrix} \cos\theta & -\sin\theta \\ \sin\theta & \cos\theta \end{pmatrix}, \quad (26)$$

where $\theta$ denotes the mixing angle. The mass eigenstates $(h_1, h_2)$ are related to the gauge eigenstates $(h, s)$ through the mixing matrix as

$$\begin{pmatrix} h \\ s \end{pmatrix} = \begin{pmatrix} \cos\theta & \sin\theta \\ -\sin\theta & \cos\theta \end{pmatrix} \begin{pmatrix} h_1 \\ h_2 \end{pmatrix}. \tag{27}$$

We emphasize that the limit $\theta \to 0$ corresponds to the SM-like limit $(h_1 \to h, h_2 \to s)$. The mass eigenvalues are obtained as

$$m^2_{h_1,h_2} = \frac{1}{2}\left(\frac{\lambda}{2}v^2 + \Lambda^2 \mp \frac{\frac{\lambda}{2}v^2 - \Lambda^2}{\cos 2\theta}\right) \tag{28}$$

$$= \frac{1}{2}\left(\frac{\lambda}{2}v^2 + \Lambda^2 \mp \sqrt{\left(\frac{\lambda}{2}v^2 - \Lambda^2\right)^2 + 4\left(\frac{\delta_2}{2}v v_S\right)^2}\right), \tag{29}$$

$$\cos 2\theta = \frac{\frac{\lambda}{2}v^2 - \Lambda^2}{m^2_{h_1} - m^2_{h_2}}. \tag{30}$$

We identify $h_1$ with the Higgs boson observed in the LHC experiments—i.e., $m_{h_1} = 125$ GeV. The $\chi$ mass is determined by the soft breaking terms $a_1$ and $b_1$ as

$$m^2_\chi = \frac{b_2}{2} - \frac{b_1}{2} + \frac{\delta_2}{4}v^2 + \frac{d_2}{4}v_S^2$$
$$= -\frac{\sqrt{2}a_1}{v_S} - b_1, \tag{31}$$

where the tadpole condition [Eq. (24)] has been used in the second equality.

For later convenience, we summarize the relations between the input and output parameters. Here, we take $\{v, m_{h_1}, m_{h_2}, \theta, a_1, v_S, m_\chi\}$ as input parameters, while the Lagrangian parameters $\{m^2, b_2, \lambda, d_2, \delta_2, b_1\}$ can be expressed in terms of the inputs. Among these, $m^2$ and $b_2$ are eliminated using the tadpole conditions in Eqs. (23) and (24):

$$m^2 = -\frac{\lambda}{2}v^2 - \frac{\delta_2}{2}v_S^2, \tag{32}$$

$$b_2 = -\frac{\delta_2}{2}v^2 - \frac{d_2}{2}v_S^2 - \sqrt{2}\frac{a_1}{v_S} - b_1. \tag{33}$$

The remaining four Lagrangian parameters are then written as

$$\lambda = \frac{2}{v^2}(m^2_{h_1}\cos^2\theta + m^2_{h_2}\sin^2\theta), \tag{34}$$

$$\delta_2 = \frac{1}{v v_S}(m^2_{h_1} - m^2_{h_2})\sin 2\theta, \tag{35}$$

$$d_2 = 2\left(\frac{m_{h_1}}{v_S}\right)^2 \sin^2\theta + 2\left(\frac{m_{h_2}}{v_S}\right)^2 \cos^2\theta + 2\sqrt{2}\frac{a_1}{v_S^3}, \tag{36}$$

$$b_1 = -m^2_\chi - \frac{\sqrt{2}}{v_S}a_1. \tag{37}$$

Theoretical constraints on the quartic couplings in the scalar potential are summarized below. To ensure that the potential is bounded from below, the following conditions must hold[2]:

$$\lambda > 0, \qquad d_2 > 0. \tag{38}$$

Furthermore, the couplings $\lambda$ and $d_2$ are constrained by perturbative unitarity [49]:

$$\lambda < \frac{16\pi}{3}, \qquad d_2 < \frac{16\pi}{3}. \tag{39}$$

In addition, requiring the eigenvalues of the mass matrix in Eq. (29) to be positive leads to the following tree-level stability condition [33]:

$$\frac{2\lambda\Lambda^2}{v_s^2} = \lambda\left(d_2 - \frac{2\sqrt{2}a_1}{v_S^3}\right) > \delta_2^2. \tag{40}$$

### B. Degenerate scalar scenario

Recent DM direct-detection experiments have provided stringent upper limits on the spin-independent DM-nucleon scattering cross section [51], which strongly constrain various Higgs-portal-type DM models. In the CxSM, the scattering process of the DM $\chi$ with nucleons is mediated by $h_1$ and $h_2$. It has been pointed out [49] that the DM-quark scattering amplitude can be significantly suppressed when the masses of $h_1$ and $h_2$ are nearly degenerate. This so-called degenerate scalar scenario naturally evades the current direct-detection bounds without introducing additional symmetry structures.

In this work, although we do not focus on the detailed DM phenomenology, the degenerate scalar region remains of particular interest, because it is simultaneously consistent with both direct-detection constraints and collider searches. The coupling of $h_1$ ($h_2$) to the SM fermions is scaled by $\cos\theta$ ($-\sin\theta$):

$$\mathcal{L}_Y = \frac{m_q}{v}\bar{q}q(h_1\cos\theta - h_2\sin\theta), \tag{41}$$

where $m_q$ denotes a mass of the quark $q$. The couplings of $h_1$ and $h_2$ to the SM gauge bosons are modified in the same

[2] For $\delta_2 < 0$, an additional condition $\lambda d_2 > \delta_2^2$ is required. In this work, we assume $\delta_2 > 0$.

way. Consequently, the partial decay widths of $h_1$ and $h_2$ into an SM final state $X$ are given by

$$\Gamma_{h_1\to XX} = \cos^2\theta \Gamma^{\rm SM}_{h\to XX}(m_{h_1}), \tag{42}$$

$$\Gamma_{h_2\to XX} = \sin^2\theta \Gamma^{\rm SM}_{h\to XX}(m_{h_2}), \tag{43}$$

where $\Gamma^{\rm SM}_{h\to XX}(m_{h_{1,2}})$ denotes the corresponding SM Higgs partial decay width evaluated at $m_{h_{1,2}}$.

When the two scalar masses are nearly degenerate ($m_{h_1} \simeq m_{h_2}$), their production and decay processes cannot be experimentally distinguished, and only the combined signal is observed:

$$\Gamma_{h_1\to XX} + \Gamma_{h_2\to XX} \simeq \Gamma^{\rm SM}_{h\to XX}(m_h). \tag{44}$$

Therefore, the total Higgs signal strength in the degenerate limit coincides with that of the SM regardless of the mixing angle $\theta$. Such a parameter region is thus phenomenologically motivated, being compatible with current Higgs measurements while remaining consistent with DM direct-detection constraints.

## IV. NUMERICAL ANALYSIS OF PBH FORMATION

Before presenting the numerical results for PBH formation, we first discuss the characteristics of the EWPT in the CxSM. The thermal evolution of the vacuum in this model proceeds from the singlet-like phase $(\langle\Phi\rangle, \langle S\rangle) = (0, v'_S)$ to the electroweak symmetry-broken phase $(v, v_S)$, where $\langle\Phi\rangle$ and $\langle S\rangle$ denote the VEVs of the SM Higgs doublet and the singlet scalar, respectively. It has been shown in previous studies that the tree-level contributions play an essential role in determining the nature of the EWPT in the CxSM [42]. In particular, the parameters $\delta_2$ and $d_2$ are the key factors that control the strength of the transition. Using the tree-level potential and thermal masses, the critical temperature $T_C$ and the Higgs VEV $v_C$ at $T_C$ can be approximately expressed as follows[3]:

$$T_C \simeq \sqrt{\frac{1}{2\Sigma_\Phi}\left(-m^2 - \frac{(v'_{SC})^2}{2}\delta_2\right)},$$
$$v_C \simeq \sqrt{\frac{2\delta_2 (v'_{SC})^2}{\lambda}\left(1 - \frac{v_{SC}}{v'_{SC}}\right)}, \tag{45}$$

where all quantities evaluated at $T_C$ are denoted with the subscript $C$. It is found that a large $\delta_2$ is favored to realize Eq. (1). $\Sigma_\Phi$ and $\Sigma_S$ represent the two-point self-energies of the fields $\Phi$ and $S$, respectively, which are related to the thermal mass corrections. (Their explicit forms can be found in Ref. [42].) On the other hand, $v'_S$ is obtained from the tadpole condition at $(0, v'_S)$ as

$$(v'_{SC})^3 + \frac{2(b_1 + b_2 + 2\Sigma_S)}{d_2} v'_{SC} + \frac{4\sqrt{2}a_1}{d_2} = 0. \tag{46}$$

When real solutions exist, $v'_S$ is scaled by $1/\sqrt{d_2}$, implying that a smaller $d_2$ is required to obtain a larger $v_C$. However, by comparing the potential energies of the two vacua and imposing the condition that the potential takes its global minimum at $(v, v_S)$, the case with $\delta_2 \gg 1$ and $d_2 \ll 1$ is excluded. An appropriate choice for realizing a strong first-order EWPT is $\delta_2 = \mathcal{O}(1)$ and $d_2 = \mathcal{O}(1)$. As seen from Eq. (35), in the Higgs-degenerate region, a small $v_S$ and a large mixing angle $\theta$ are favored in order to realize a moderately large $\delta_2$. Moreover, as seen from Eq. (36), $d_2$ depends inversely on $v_S^2$ (and partly on $v_S^3$ through the $a_1$ term). With the small value of $v_S$ required to realize an appropriate $\delta_2$, the value of $d_2$ becomes highly sensitive to small variations in $a_1$ and $m_{h_2}$.

Figure 1 illustrates the parameter dependence of PBH formation in the $(\delta_2, d_2)$ plane. Here, we take $v_S = 0.6$ GeV and $\theta = \pi/4$ as representative values to realize appropriate magnitudes of $\delta_2$ and $d_2$ in the degenerate scalar scenario. The red band indicates the region where the PBH fraction satisfies $10^{-4} \le f_{\rm PBH} \le 1$, corresponding to efficient PBH formation. This narrow red band, which almost appears as a single line, indicates that $f_{\rm PBH}$ is extremely sensitive to small variations in the parameters $\delta_2$ and $d_2$. In fact, as shown in Eqs. (17) and (18), the PBH fraction $f_{\rm PBH}$ is proportional to the probability of PBH formation $P(t_N)$, which is itself proportional to the decay rate $\Gamma$ in Eq. (2). Since $\Gamma$ is approximately given by $e^{-S_3/T}$, we eventually obtain $f_{\rm PBH} \propto e^{e^{-S_3/T}}$. Therefore, PBH production is highly sensitive to the details of $S_3/T$. The upward slope reflects that a stronger first-order EWPT tends to occur for larger $\delta_2$ and smaller $d_2$, and that efficient PBH formation is realized only in a limited region of this parameter space. The color map shows the deviation of the Higgs triple coupling from the SM value, $g_{hhh}^{\rm SM} = 3m_h^2/v$, defined as

$$R \equiv \frac{g_{h_1h_1h_1} - g_{hhh}^{\rm SM}}{g_{hhh}^{\rm SM}} \times 100[\%] \quad \text{with} \quad \mathcal{L} = g_{h_1h_1h_1} h_1^3. \tag{47}$$

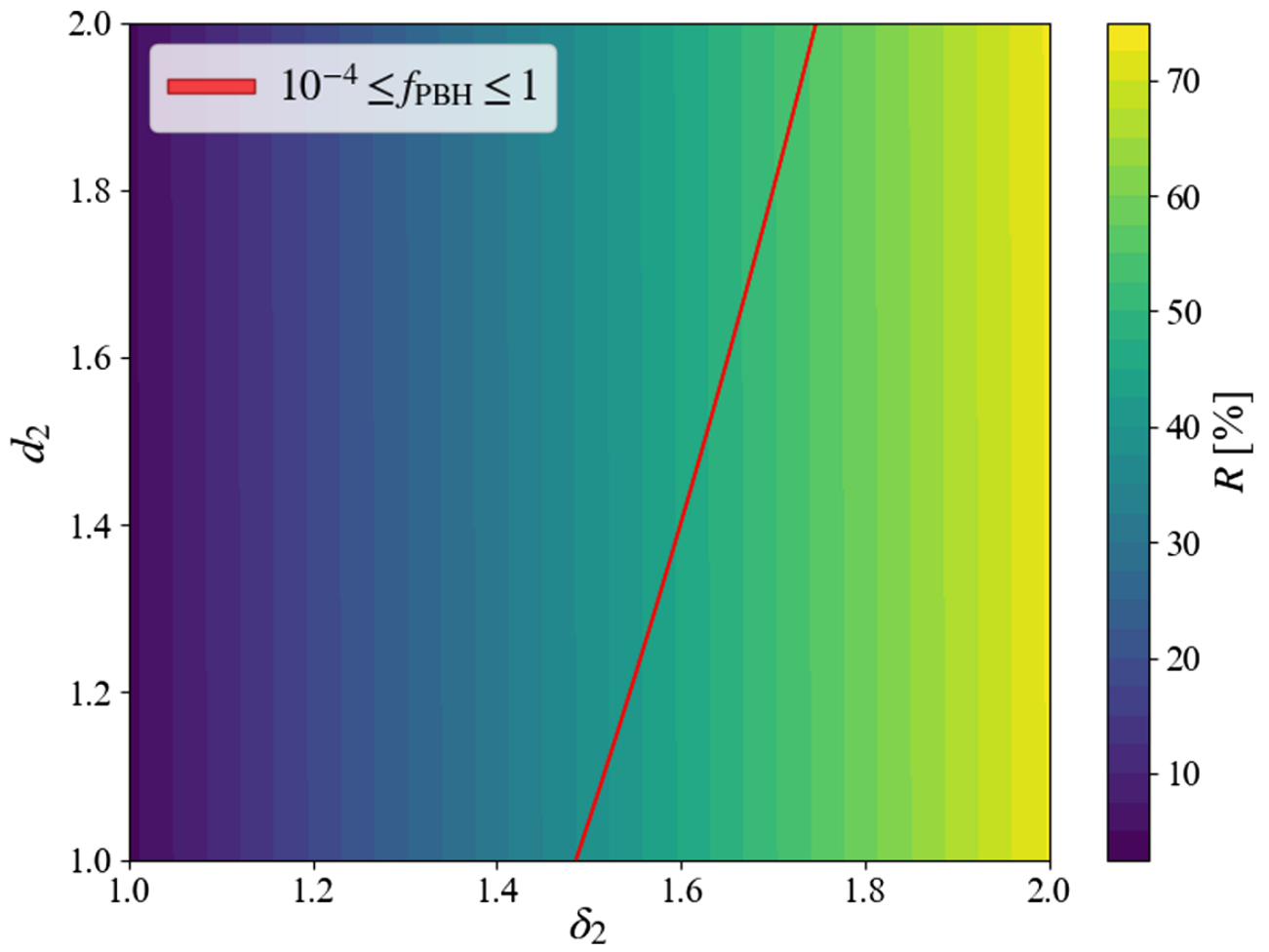

FIG. 1. Parameter dependence of the PBH fraction in the $(\delta_2, d_2)$ plane. Here we take $v_S = 0.6$ GeV and $\theta = \pi/4$ as representative values to realize appropriate magnitudes of $\delta_2$ and $d_2$ in the degenerate scalar scenario. The red line represents the region where $10^{-4} \le f_{\rm PBH} \le 1$. The color map shows the deviation of the Higgs triple coupling from the SM value defined as Eq. (47), which serves as an indicator of the EWPT strength.

[3]Although the cubic terms in the fields arising from thermal boson loops serve as the source of the EWPT, we focus on the tree-level potential supplemented by the thermal masses (quadratic terms in the fields), which dominate at high temperature, in order to illustrate that the tree-level potential plays an essential role in forming the potential barrier between vacua. In the numerical analysis presented later, we employ the full one-loop effective potential.

TABLE I. Benchmark points used in the analysis. The left columns show the input parameters, while the right columns summarize the derived parameters $\delta_2$ and $d_2$ computed from Eqs. (35) and (36).

| | Input parameters | | | | | | | Derived parameters | |
|---|---|---|---|---|---|---|---|---|---|
| Benchmark | $v$ [GeV] | $m_{h_1}$ [GeV] | $m_{h_2}$ [GeV] | $\theta$ [rad] | $a_1$ [GeV$^3$] | $v_S$ [GeV] | $m_\chi$ [GeV] | $\delta_2$ | $d_2$ |
| BP1 | 246.22 | 125 | 124 | $\pi/4$ | −6576.1720 | 0.6 | 62.5 | 1.6855 | 1.7442 |
| BP2 | 246.22 | 125 | 124 | $\pi/4$ | −6576.1723 | 0.6 | 62.5 | 1.6855 | 1.7402 |
| BP3 | 246.22 | 125 | 124 | $\pi/4$ | −6576.1724 | 0.6 | 62.5 | 1.6855 | 1.7389 |
| BP4 | 246.22 | 125 | 123.999995 | $\pi/4$ | −6576.1720 | 0.6 | 62.5 | 1.6855 | 1.7407 |
| BP5 | 246.22 | 125 | 123.999993 | $\pi/4$ | −6576.1720 | 0.6 | 62.5 | 1.6855 | 1.7393 |

Although $R$ is not directly related to PBH formation, it provides a useful indicator of the strength of the EWPT. The triple coupling of the SM-like state can be written as

$$g_{h_1h_1h_1} = \frac{3}{2}[\lambda v \cos^3\theta + \delta_2 v_S \cos^2\theta \sin\theta + \delta_2 v \cos\theta \sin^2\theta + d_2 v_S \sin^3\theta]. \tag{48}$$

Among these terms, the term proportional to $d_2$ is suppressed by the small singlet VEV $v_S$, which we take to be small in the Higgs-degenerate setup. As a result, the contribution of the $d_2$ term to $g_{h_1h_1h_1}$ becomes subdominant, and the dependence of the triple coupling on $d_2$ is hardly visible in our parameter scan.

Next, as an illustrative example, we fix the value of $\delta_2$ and slightly vary $d_2$. For clarity, instead of using the parameters $(\delta_2, d_2)$, we reexpress the parameter space in terms of the physical quantities $(m_{h_2}, a_1)$, which are more directly related to observable properties of the model. This representation allows us to visualize how a small variation in $m_{h_2}$ or $a_1$ can lead to a significant change in $S_3/T$, reflecting the strong sensitivity of PBH formation to these parameters.

The benchmark points are summarized in Table I. We treat $(v, m_{h_1}, m_{h_2}, \theta, a_1, v_S, m_\chi)$ as input parameters, while $(\delta_2, d_2)$ are derived quantities calculated from Eqs. (35) and (36). BP1 serves as the reference point, where the value of $\delta_2 = 1.6855$ is chosen so that the model parameters reproduce the setup studied in Refs. [42,43]. Reference [49] notes that the mass difference $|m_{h_1} - m_{h_2}| \lesssim 3$ GeV has not been conclusively ruled out by LHC experiments [69]. In addition, at the ILC 250 [70], the recoil mass technique may allow for the discrimination of a 1 GeV mass difference, but resolving mass differences smaller than 1 GeV would require even higher precision. Thus, we focus on the Higgs mass differences of around 1 GeV. The DM mass $m_\chi$ is fixed at 62.5 GeV as an example, since it has a negligible impact on the dynamics of the EWPT and PBH formation. In BP2 and BP3, the parameter $a_1$ is varied with all other inputs fixed, whereas in BP4 and BP5, $m_{h_2}$ is slightly varied instead. The strong sensitivity of $d_2$ to these small variations originates from the small value of $v_S$ adopted in our setup. As seen from Eq. (36), $d_2$ scales as $1/v_S^2$ or partly $1/v_S^3$, so that even a tiny change in $a_1$ or $m_{h_2}$ leads to a shift in $d_2$, as shown in Table I. In contrast, the dependence of $\delta_2$ on $m_{h_2}$ is comparatively weak. These benchmark points are therefore suitable for probing how small variations in $a_1$ or $m_{h_2}$ can influence $S_3/T$, and consequently, the PBH abundance.

TABLE II. Parameters characterizing the EWPT for each benchmark point.

| | EWPT parameters | | | |
|---|---|---|---|---|
| Benchmark | $v_C$ [GeV] | $T_C$ [GeV] | $v_C/T_C$ | $T_N$ [GeV] |
| BP1 | 201.05 | 105.41 | 1.9073 | 56.442 |
| BP2 | 201.20 | 105.31 | 1.9106 | 54.751 |
| BP3 | 201.25 | 105.28 | 1.9116 | 54.094 |
| BP4 | 201.18 | 105.32 | 1.9102 | 54.965 |
| BP5 | 201.23 | 105.28 | 1.9113 | 54.283 |

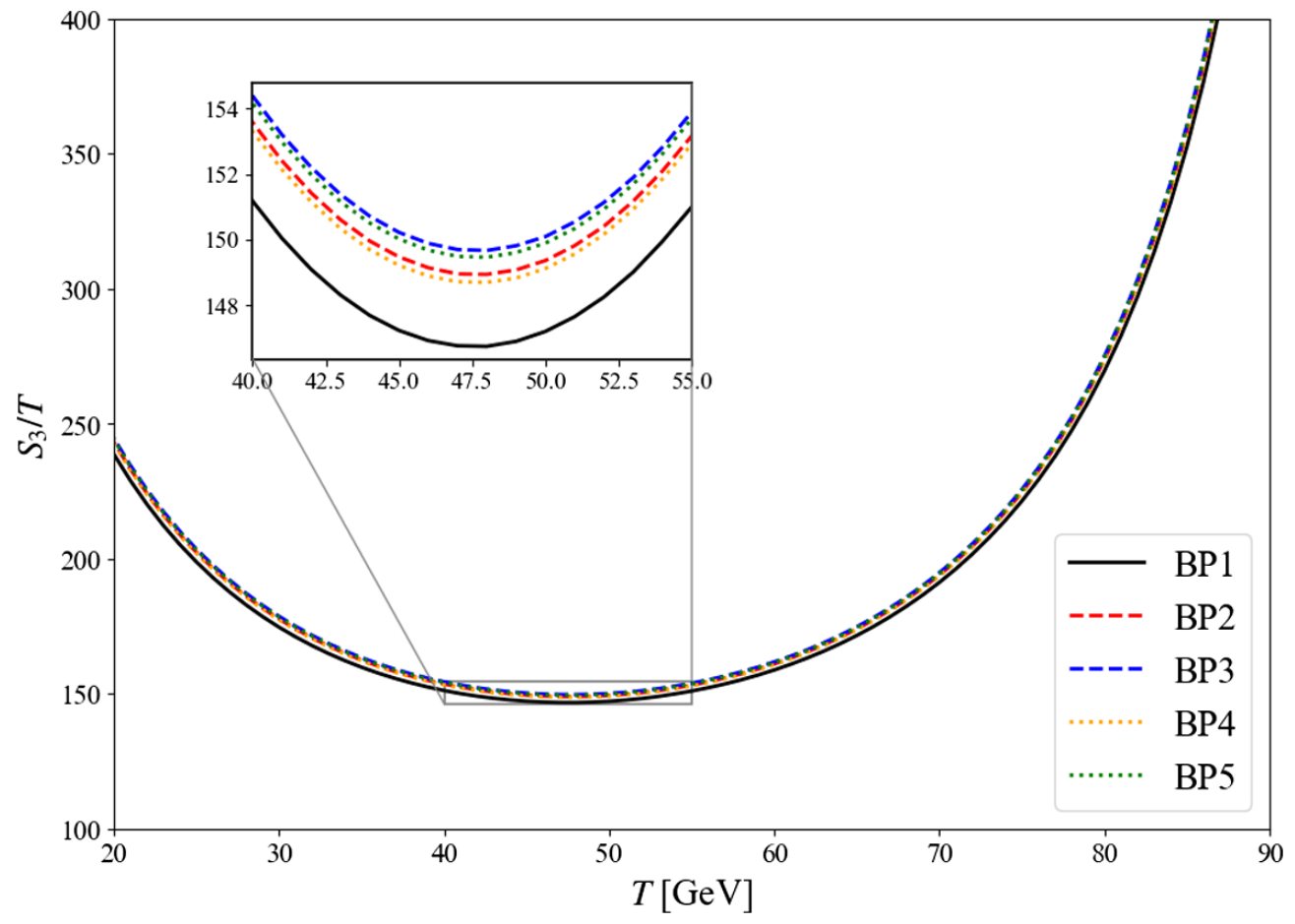

FIG. 2. Temperature dependence of $S_3/T$, which characterizes the strength of the EWPT. The inset shows an enlarged view around the minimum of $S_3/T$.

The parameters relevant to the EWPT are summarized in Table II. $v_C$, $T_C$, and $T_N$, defined in Eq. (3), are somewhat sensitive to such small parameter changes. A strong first-order EWPT with $v_C/T_C > 1$ is realized at all benchmark points. It is also seen that a smaller $d_2$ generally results in a slightly stronger first-order EWPT, as reflected in the larger values of $v_C/T_C$. Furthermore, Fig. 2 shows the temperature dependence of $S_3/T$, which reflects the strength of the EWPT. All benchmark points exhibit $U$-shaped $S_3/T$ curves, as typically seen in a strong first-order EWPT. The inset highlights the region near the minimum of $S_3/T$. Benchmark points corresponding to a stronger first-order EWPT show slightly larger values of $S_3/T$.

The results of the PBH analysis are summarized in Table III and visualized in Fig. 3. The quantities shown in Table III are defined in Sec. II. $T_d$ denotes the reference temperature of the delayed-decayed region; $T_{\mathrm{PBH}}$ is the PBH formation temperature; $f_{\mathrm{PBH}}$ in Eq. (17) represents the present PBH fraction; and $M_{\mathrm{PBH}}$ in Eq. (19) is the PBH mass, normalized by the solar mass $M_\odot$. Figure 3 shows the predicted PBH abundance $f_{\mathrm{PBH}}$ as a function of the PBH mass normalized by the solar mass ($M_{\mathrm{PBH}}/M_\odot$), together with the existing microlensing constraints from Subaru HSC [54], the long-term OGLE Large Magellanic Cloud survey (OGLE-LMC) [55], and the OGLE high-cadence Magellanic Clouds survey (OGLE-HC) [56]. The red points correspond to the benchmark points BP2–BP5, while BP1 is not visible, since its PBH fraction is too small to appear in the plot. As seen from Table III, the predicted PBH masses are all around $M_{\mathrm{PBH}} \sim 10^{-5} M_\odot$, reflecting the similar nucleation temperatures among the benchmark points.

TABLE III. Parameters related to PBH formation for each benchmark point. Here, $T_d$ is the reference temperature of the delayed-decayed region; $T_{\mathrm{PBH}}$ is the temperature at which PBH is produced; $f_{\mathrm{PBH}}$ denotes the present PBH fraction; and $M_{\mathrm{PBH}}$ is the PBH mass, which is normalized by the solar mass $M_\odot$.

| PBH-related parameters | | | | |
|---|---|---|---|---|
| Benchmark | $T_d$ [GeV] | $T_{\mathrm{PBH}}$ [GeV] | $f_{\mathrm{PBH}}$ | $M_{\mathrm{PBH}}/M_\odot$ |
| BP1 | 44.292 | 29.375 | $3.01 \times 10^{-214}$ | $6.95 \times 10^{-6}$ |
| BP2 | 46.581 | 21.364 | $2.09 \times 10^{-12}$ | $1.03 \times 10^{-5}$ |
| BP3 | 47.464 | 15.567 | 5.68 | $1.47 \times 10^{-5}$ |
| BP4 | 46.287 | 21.960 | $3.55 \times 10^{-21}$ | $1.00 \times 10^{-5}$ |
| BP5 | 47.208 | 17.647 | $3.68 \times 10^{-2}$ | $1.27 \times 10^{-5}$ |

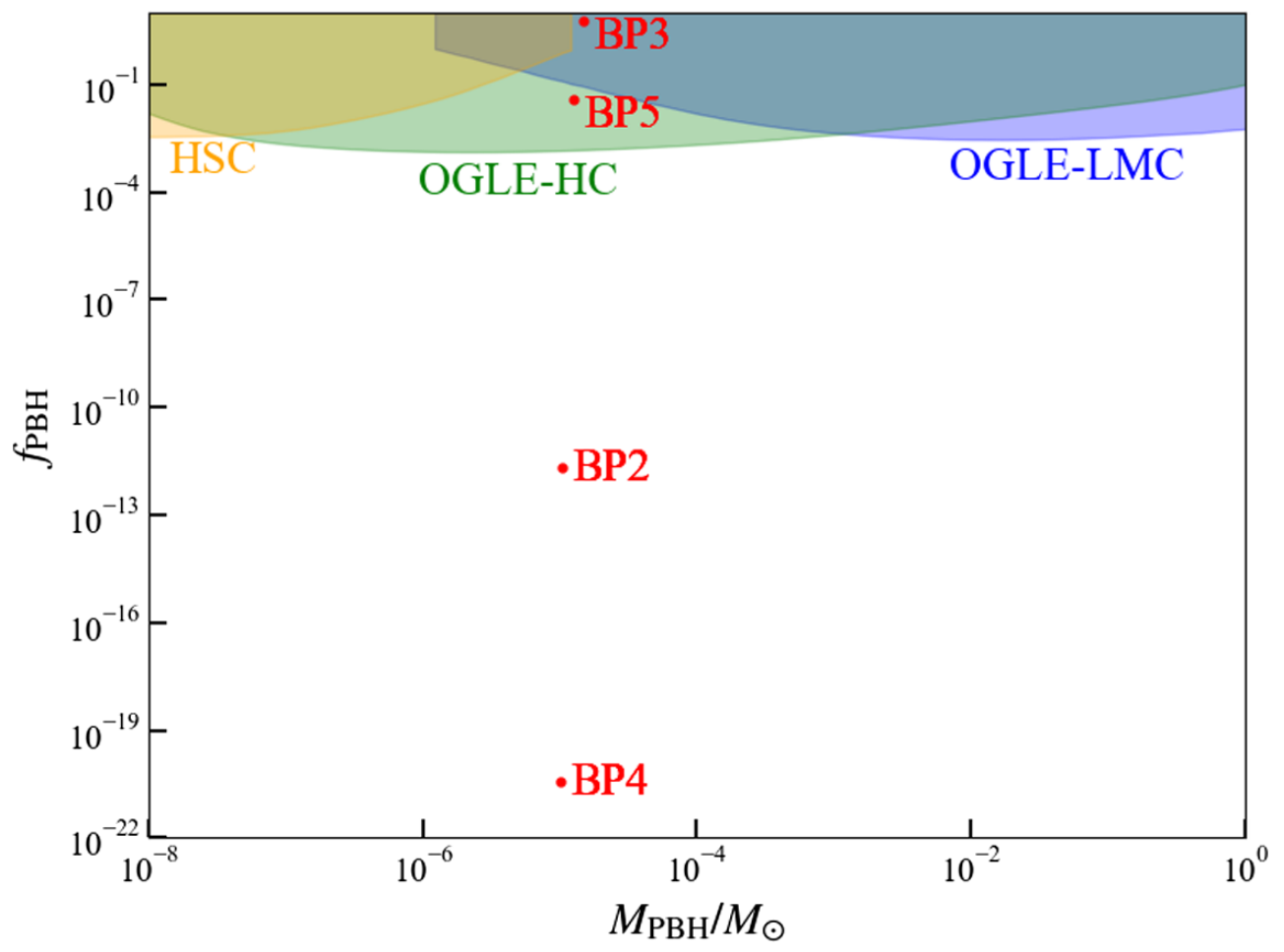

FIG. 3. Predicted PBH abundance $f_{\mathrm{PBH}}$ as a function of the PBH mass, normalized by the solar mass ($M_{\mathrm{PBH}}/M_\odot$). The colored shaded regions represent current microlensing constraints: Subaru HSC [54] (HSC, orange), the long-term OGLE Large Magellanic Cloud survey (OGLE-LMC, blue) [55], and the OGLE high-cadence Magellanic Clouds survey (OGLE-HC, green) [56]. The red points indicate the benchmark scenarios BP2–BP5, while BP1 is omitted because its predicted PBH fraction is too small to appear in the plot.

## V. GRAVITATIONAL WAVE IN SYNERGY WITH LEPTON COLLIDER SIGNATURES

In addition to PBH formation, the first-order EWPT can generate observable GWs, providing an independent probe of the underlying dynamics of the phase transition. Together with collider measurements, such as the deviation of the Higgs triple coupling from its SM value, GWs offer a complementary way to explore the electroweak-scale physics responsible for the phase transition. In this section, we study the GW signal predicted in the present model and discuss its detectability in future space-based interferometers, as well as its correlation with collider observables. Since the favored parameter region for the PBH production could lead to obvious deviation of the Higgs triple coupling as shown in Eq. (47), future lepton colliders, such as CEPC, FCC-ee, and ILC could be able to test this scenario through loop effects.

In our analysis, we employ the formalism of Ref. [71] to evaluate the GW signals. One of the key quantities that characterize the dynamics of the phase transition is the strength parameter $\alpha$, defined as the ratio of the released vacuum energy density to the radiation energy density of the plasma:

$$\alpha \equiv \frac{\Delta\rho_v}{\rho_r} = \frac{\left[\Delta V - T\frac{d\Delta V}{dT}\right]_{T=T_*}}{\frac{\pi^2}{30} g_* T_*^4}, \tag{49}$$

where $\Delta V$ is the potential energy difference between the false and true vacua, $T_*$ denotes the characteristic temperature of the transition, and $g_*$ is the effective number of relativistic degrees of freedom at $T_*$. Here, we adopt $g_* = 108.75$. A larger value of $\alpha$ corresponds to a stronger first-order transition.

Another important parameter characterizing the transition is the characteristic length scale, which corresponds to the typical distance between nucleated bubbles at the time of the transition. This scale is expected to carry most of the released energy, and therefore plays a crucial role in determining the GWs generated by bulk fluid motion and bubble collisions. The mean bubble separation is usually adopted as this characteristic length scale. Following Ref. [71], the mean bubble separation, or the effective bubble radius $R_*$, is given by

$$R_* = n_b^{-1/3}, \qquad n_b = \int_{t_C}^{t} dt' \frac{a(t')^3}{a(t)^3} \Gamma(t') F(t'), \tag{50}$$

where $n_b$ denotes the number density of bubbles nucleated up to time $t$.

Sources of GWs arise from bubble collisions [72–77], sound waves [78–81], and turbulence induced by the percolation process [82–87]. The total GW spectrum is the sum of these three contributions:

$$h_{100}^2 \Omega_{\rm GW}(f) = h_{100}^2 \Omega_{\rm co}(f) + h_{100}^2 \Omega_{\rm sw}(f) + h_{100}^2 \Omega_{\rm turb}(f). \tag{51}$$

Here, $h_{100} = H_0/(100\ {\rm km\,s^{-1}\,Mpc^{-1}}) = 0.67$ is the dimensionless Hubble parameter from the Planck data, with $H_0$ being the Hubble constant today [88]. We compute the GW spectrum using the $R_*$-based fitting formulas of Ref. [71]. The spectral contributions from bubble collisions, sound waves, and turbulence are given, respectively, by

$$\begin{aligned}
h_{100}^2 \Omega_{\rm co}(f) &\simeq 1.67 \times 10^{-5} \left(\frac{H_* R_*}{(8\pi)^{1/3}}\right)^2 \left(\frac{\kappa_{\rm co}\alpha}{1+\alpha}\right)^2 \left(\frac{100}{g_*}\right)^{1/3} \frac{0.11 v_w}{0.42 + v_w^2} \frac{3.8 (f/f_{\rm co})^{2.8}}{1 + 2.8 (f/f_{\rm co})^{3.8}}, \\
h_{100}^2 \Omega_{\rm sw}(f) &\simeq 1.64 \times 10^{-6} (H_* \tau_{\rm sw})(H_* R_*) \left(\frac{\kappa_v \alpha}{1+\alpha}\right)^2 \left(\frac{100}{g_*}\right)^{1/3} \left(\frac{f}{f_{\rm sw}}\right)^3 \left[\frac{7}{4 + 3(f/f_{\rm sw})^2}\right]^{7/2}, \\
h_{100}^2 \Omega_{\rm turb}(f) &\simeq 1.14 \times 10^{-4} (H_* R_*) \left(\frac{\kappa_{\rm turb}\alpha}{1+\alpha}\right)^{3/2} \left(\frac{100}{g_*}\right)^{1/3} \frac{(f/f_{\rm turb})^3}{(1 + f/f_{\rm turb})^{11/3} (1 + 8\pi f/H_*)}.
\end{aligned} \tag{52}$$

Here, $H_*$ and $R_*$ are evaluated at the percolation temperature $T_* = T_p$, defined by $F(T_p) = 0.7$. The duration of the sound-wave source is

$$\tau_{\rm sw} = \min\left[\frac{1}{H_*}, \frac{R_*}{\bar{U}_f}\right], \tag{53}$$

where the root-mean-square fluid velocity is approximated as $\bar{U}_f^2 \simeq \frac{3}{4}\frac{\kappa_v \alpha}{1+\alpha}$.

For $v_w \simeq 1$, the efficiency factors are approximated by [71,89,90]

$$\begin{aligned}
\kappa_{\rm co} &\simeq \frac{1}{1 + 0.715\alpha}\left(0.715\alpha + \frac{4}{27}\sqrt{\frac{3\alpha}{2}}\right), \\
\kappa_v &\simeq \frac{\alpha}{0.73 + 0.083\sqrt{\alpha} + \alpha}, \qquad \kappa_{\rm turb} \simeq (0.05 - 0.1)\kappa_v.
\end{aligned} \tag{54}$$

In our numerical analysis, we set $v_w = 0.95$ and $\kappa_{\rm turb} = 0.1\kappa_v$ for illustration. The corresponding peak frequencies are

$$\begin{aligned}
f_{\rm co} &\simeq 1.65 \times 10^{-5}\ {\rm Hz} \frac{(8\pi)^{1/3}}{H_* R_*} \frac{0.62 v_w}{1.8 - 0.1 v_w + v_w^2} \left(\frac{T_*}{100\ {\rm GeV}}\right)\left(\frac{g_*}{100}\right)^{1/6}, \\
f_{\rm sw} &\simeq 2.6 \times 10^{-5}\ {\rm Hz} \frac{1}{H_* R_*} \left(\frac{T_*}{100\ {\rm GeV}}\right)\left(\frac{g_*}{100}\right)^{1/6}, \\
f_{\rm turb} &\simeq 7.91 \times 10^{-5}\ {\rm Hz} \frac{1}{H_* R_*} \left(\frac{T_*}{100\ {\rm GeV}}\right)\left(\frac{g_*}{100}\right)^{1/6}.
\end{aligned} \tag{55}$$

To quantitatively assess the detectability of the predicted GWs, we compute the signal-to-noise ratio (SNR) for each benchmark point. The SNR is evaluated as follows:

$$\mathrm{SNR} = \sqrt{\mathcal{T} \int_{f_{\min}}^{f_{\max}} df \left( \frac{h_{100}^2 \Omega_{\mathrm{GW}}(f)}{h_{100}^2 \Omega_{\mathrm{sens}}(f)} \right)^2}, \tag{56}$$

where $\mathcal{T}$ denotes the observation time, and $h_{100}^2 \Omega_{\mathrm{sens}}(f)$ represents the detector sensitivity curve [91]. In our analysis, we adopt $\mathcal{T} \simeq 1.0 \times 10^8$ s for both LISA and TianQin, corresponding to approximately 4 yr of observation time. A GW signal is considered detectable if $\mathrm{SNR} > \mathrm{SNR}_{\mathrm{thr}}$, where we set the threshold value $\mathrm{SNR}_{\mathrm{thr}} = 5$.

The GW spectra in this work are computed using the semianalytic prescription developed in Ref. [71], which provides a practical framework for estimating GW signals from strongly supercooled first-order phase transitions based on the mean bubble separation. More refined treatments of GW production in the supercooling regime have also been discussed in the literature, such as those presented in Refs. [92,93]. In particular, Ref. [93] incorporates the distribution of bubble radii and collision times through a radius-dependent efficiency factor, showing that these effects mainly modify the overall normalization of the GW spectrum while leaving its spectral shape largely unchanged. As a result, although quantitative differences can arise among different prescriptions due to scheme dependence, they lead to consistent qualitative trends for the GW signal as a function of the phase transition parameters. Since the focus of the present work is on the correlations between PBH formation, GW observables, and collider signatures, the use of the prescription of Ref. [71] is sufficient for the purposes of our analysis.

Table IV summarizes the key parameters characterizing the GW and collider phenomenology of the benchmark points. The quantities $T_p$, $\alpha_p$, and $(R_*)_p$ represent the percolation temperature, the strength of the phase transition, and the mean bubble separation at $T_p$, respectively. Larger $\alpha_p$ and $R_*$ correspond to a stronger and more prolonged phase transition, which in turn leads to more energetic GW production. The deviation of the Higgs triple coupling from the SM value $R$ is around 50% for all benchmark points. At the one-loop level, this Higgs triple coupling can increase the SM $Zh$ cross section ($\delta\sigma_{Zh} \equiv \sigma_{Zh}/\sigma_{Zh}^{\mathrm{SM}} - 1$) about $\delta\sigma_{Zh} = 0.82\%$ at 240 GeV CEPC [94,95]. Under the CEPC TDR nominal operating conditions, the cross section is projected to be measured with a relative precision of 0.26% [96,97]. Recent advances in AI-assisted reconstruction and analysis suggest that this precision could be significantly improved, potentially reaching 0.1% in an aggressive scenario [96,97]. One can see that CEPC has the ability to test this scenario with high precision. Similar discussions could also be performed at ILC and FCC-ee. For ILC, the expected precision of the $Zh$ cross section is about 0.4% [98]. For FCC-ee, the expected precision is about 0.1% [99,100]. Current Higgs data at LHC require the Higgs triple coupling modification to be within the range $(-0.4, 6.3)$. At the High-Luminosity LHC (HL-LHC), with a projected integrated luminosity of 3 $\mathrm{ab}^{-1}$ of high-quality data per experiment, the ATLAS and CMS Collaborations anticipate determining the Higgs triple coupling with a precision better than 30% [101]. Since this scenario can modify the Higgs triple coupling around 50%, CEPC, ILC, FCC-ee, and HL-LHC are able to test this scenario. We also list the SNRs for LISA and TianQin, computed under the assumption of four years of observation. All benchmark points yield $\mathrm{SNR} > 5$ for LISA and comparable values for TianQin, implying that the predicted GW signals would be well within the reach of both detectors. Overall, this table illustrates that in the CxSM, a strong first-order EWPT leading to PBH formation can simultaneously give rise to sizable deviations in the Higgs self-coupling and detectable GWs, providing mutually complementary probes of the same underlying dynamics.

Figure 4 visualizes the GW spectra corresponding to the SNR results summarized in Table IV. The colored curves represent the predicted spectra for the benchmark points, while the shaded regions indicate the expected sensitivity ranges of future space-based detectors LISA (red) and TianQin (blue). As expected from the SNR values, the overall amplitude of the GW spectra increases with the strength of the first-order EWPT. This trend reflects the correlation between a strongly supercooled transition, efficient PBH formation, and enhanced GW production. All benchmark points yield GW spectra within the sensitivity bands of both LISA and TianQin, demonstrating that the parameter space leading to PBH formation can be simultaneously tested by upcoming GW observations.

TABLE IV. Parameters relevant to the GW and collider phenomenology. Here, $T_p$, $\alpha_p$, and $(R_*)_p$ denote the percolation temperature, the phase transition strength, and the mean bubble separation at $T_p$, respectively. $R$ is the relative deviation of the Higgs triple coupling from the SM value defined in Eq. (47) and the predicted modification of the $Zh$ production cross section, $\delta\sigma_{Zh} \equiv \sigma_{Zh}/\sigma_{Zh}^{\mathrm{SM}} - 1$, which reflects the loop-level effect of the enhanced Higgs self-coupling. SNRs for LISA and TianQin are computed assuming an observation time of $\mathcal{T} \simeq 1.0 \times 10^8$ s. The line styles correspond to those used in Fig. 4.

| GW and collider-related parameters | | | | | | | | |
|---|---|---|---|---|---|---|---|---|
| Benchmark | $T_p$ [GeV] | $\alpha_p$ | $(R_*)_p$ [GeV$^{-1}$] | $R$ [%] | $\delta\sigma_{Zh}$ | SNR@LISA | SNR@TianQin | GW |
| BP1 | 43.581 | 0.62256 | $7.07 \times 10^{13}$ | 51.25 | 0.82% | 93.35 | 16.81 | Solid black line |
| BP2 | 38.047 | 1.0108 | $1.61 \times 10^{14}$ | 51.25 | 0.82% | 59.15 | 9.32 | Dashed red line |
| BP3 | 35.287 | 1.3321 | $2.22 \times 10^{14}$ | 51.25 | 0.82% | 47.48 | 6.48 | Dashed blue line |
| BP4 | 38.848 | 0.93730 | $1.46 \times 10^{14}$ | 51.25 | 0.82% | 63.02 | 10.32 | Dotted orange line |
| BP5 | 36.121 | 1.2222 | $2.03 \times 10^{14}$ | 51.25 | 0.82% | 50.82 | 7.25 | Dotted green line |

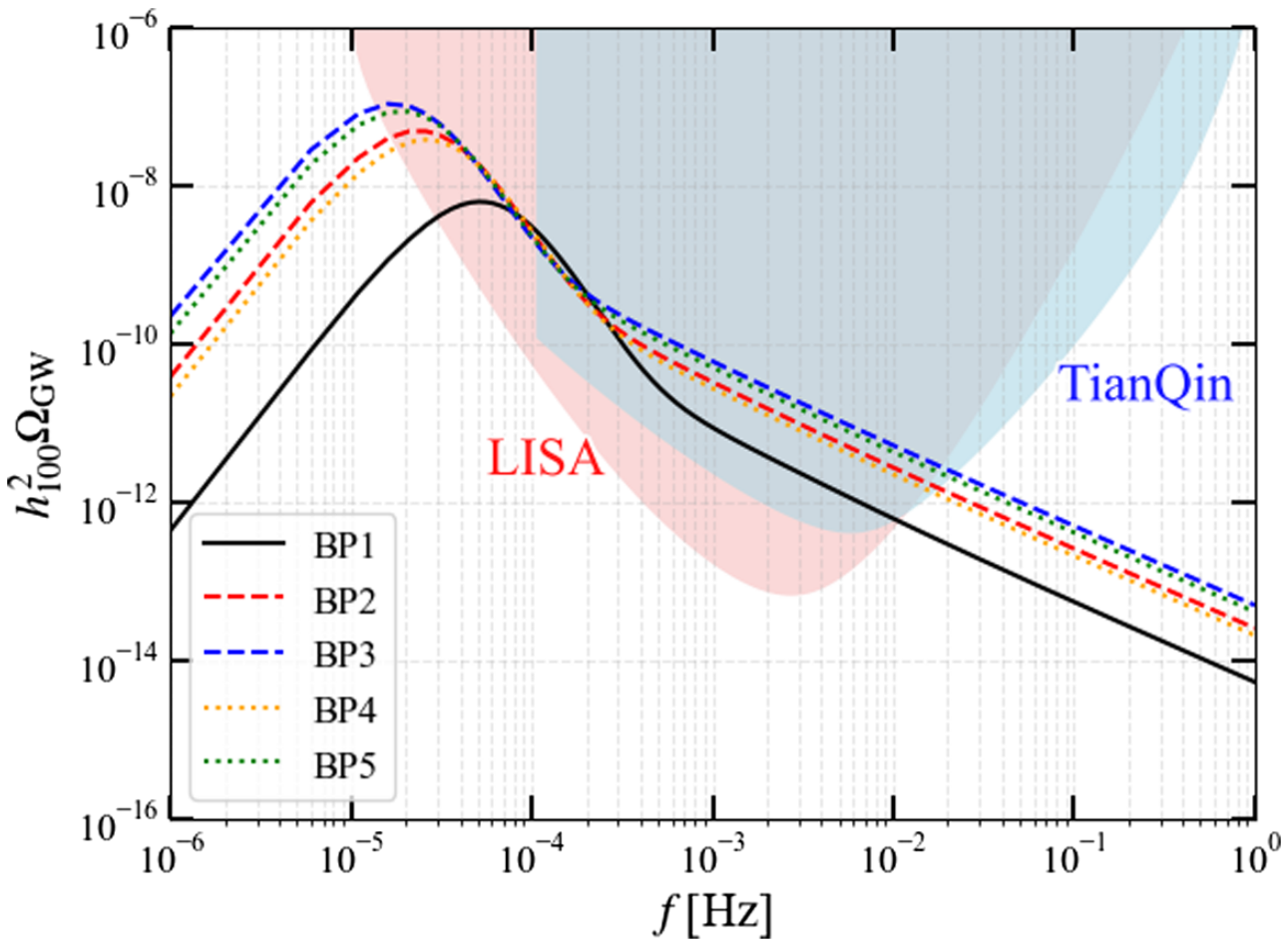

FIG. 4. The predicted GW spectra for the benchmark points BP1–BP5. The shaded regions indicate the sensitivity curves corresponding to SNR $= 5$ of future space-based GW detectors, LISA [102–104] (red) and TianQin [105,106] (blue). GW signals above these curves are detectable with SNR $> 5$.

Beyond these specific benchmark realizations, the points shown above are chosen as representative examples to illustrate the multimessenger implications of a strong first-order EWPT in the CxSM. More generally, the underlying correlations among different observables are driven by the properties of the scalar potential. In particular, a stronger EWPT typically leads to a larger released vacuum energy, resulting in a stronger stochastic GW signal, while simultaneously inducing a larger deviation in the Higgs triple coupling. PBH formation through delayed vacuum decay is even more sensitive to the EWPT dynamics, as the PBH abundance depends exponentially on the nucleation rate and can vary dramatically across the allowed parameter space. Therefore, while the benchmark points provide concrete realizations where PBH formation, GW signals, and collider observables coexist, the complementarity of these probes is a generic feature associated with a strong first-order EWPT.

## VI. CONCLUSIONS AND DISCUSSIONS

In this work, we have investigated the possibility of PBH formation induced by a first-order EWPT in a realistic model—namely, the CxSM. This model contains a scalar DM candidate and two Higgs bosons, and it is known to be compatible with both DM direct-detection constraints and Higgs search results when the two Higgs states are nearly degenerate in mass.

We have shown that the strength of the EWPT and the resulting PBH fraction are strongly dependent on the quartic couplings $\delta_2$ and $d_2$ in the scalar potential. A strong first-order EWPT, required for efficient PBH formation, is realized for $\delta_2 = \mathcal{O}(1)$ and $d_2 = \mathcal{O}(1)$, which are consistent with the tree-level vacuum stability conditions. In addition, due to the Higgs mass degeneracy, a small singlet VEV $v_S$ and a large mixing angle $\theta$ are found to be necessary for realizing a strong first-order EWPT.

To further explore the sensitivity of PBH formation, we have analyzed benchmark points that differ only by tiny variations in $a_1$ and $m_{h_2}$. Even such small shifts induce large changes in the Euclidean action $S_3/T$, and hence in the PBH fraction, reflecting the double-exponential sensitivity $f_{\rm PBH} \propto e^{e^{-S_3/T}}$. In practice, a larger $f_{\rm PBH}$ is obtained for stronger first-order EWPT, while milder transitions yield negligible fractions. Depending on the benchmark points, we found fractions of phenomenological interest that remain consistent with current microlensing bounds. This sensitivity also implies that theoretical and numerical choices—e.g., the thermal resummation scheme—can lead to appreciable differences in the predicted $f_{\rm PBH}$; a systematic assessment of such scheme dependence is left for future work. Related studies can be found in Ref. [107].

We have also analyzed the GW signals generated by the first-order EWPT using the $R_*$ formalism, which accounts for the contributions from bubble collisions, sound waves, and turbulence. The resulting spectra reflect the strength of the transition, with stronger first-order EWPT producing larger GW amplitudes. All benchmark points yield SNRs exceeding 10 for LISA and comparable sensitivities for TianQin, indicating that the predicted signals are within the reach of future space-based detectors. In the same parameter region, the deviation of the Higgs triple coupling from its SM value reaches about 50% (testable at CEPC, ILC, FCC-ee, and HL-LHC) and therefore modifies the $e^+e^- \to Zh$ cross section about 0.82% at one-loop level at 240 GeV CEPC. These results suggest that both GW and collider measurements can provide complementary probes of this realistic PBH model from EWPT.

In summary, we have demonstrated that in the degenerate scalar scenario of the CxSM, a strong first-order EWPT can simultaneously lead to PBH formation, observable GW signals, and measurable deviations in the Higgs self-coupling. These results reveal the multimessenger nature of electroweak-scale physics, showing that PBH, GW, and collider observations together provide powerful and complementary means of probing their common origin from a first-order EWPT in the early Universe.

## ACKNOWLEDGMENTS

We are grateful to Masanori Tanaka and Tomo Takahashi for helpful discussions on the PBH calculations, to Siyu Jiang for valuable advice on the GW computations, and to Chen Zhou for discussions on the future sensitivity of Higgs triple coupling at HL-LHC. This work is supported by the National Natural Science Foundation of China (NNSFC) Grants No. 12475111, No. 12205387, and the Fundamental Research Funds for the Central Universities, Sun Yat-sen University.

## DATA AVAILABILITY

The data that support the findings of this article are not publicly available upon publication because it is not technically feasible and/or the cost of preparing, depositing, and hosting the data would be prohibitive within the terms of this research project. The data are available from the authors upon reasonable request.